\documentclass[journal]{IEEEtran}
\usepackage{booktabs}

\usepackage{amssymb}
\usepackage{amsbsy}
\usepackage{amsmath}
\usepackage{bm}
\usepackage{verbatim}
\usepackage{cite}
\usepackage{mathrsfs}
\usepackage{amsfonts}
\usepackage{graphicx}
\usepackage[tight,footnotesize]{subfigure}
\usepackage[10pt]{moresize}
\usepackage{array}
\usepackage{color}
\usepackage{epsfig}
\usepackage{stfloats}
\usepackage{balance}

\usepackage{caption}

\usepackage[noend]{algpseudocode}

\usepackage{algorithmicx,algorithm}

\usepackage{cite}

\usepackage{setspace}

\ifCLASSINFOpdf
\else
\fi
\begin{document}

\title{Improving Secrecy with Nearly Collinear Main and Wiretap Channels via a Cooperative Jamming Relay }


\author{Shuai~Han,~\IEEEmembership{Senior~Member,~IEEE,}
        Sai~Xu,~\IEEEmembership{Student~Member,~IEEE,}
        Weixiao~Meng,~\IEEEmembership{Senior~Member,~IEEE,}
        and~Cheng~Li,~\IEEEmembership{Senior~Member,~IEEE}% <-this % stops a space
\thanks{Sai~Xu, Shuai~Han and Weixiao~Meng are with the Communications Research Center, Harbin Institute of Technology, China. (e-mail:
fenicexusai@163.com; hanshuai@hit.edu.cn; wxmeng@hit.edu.cn).}% <-this % stops a space
\thanks{Cheng~Li is with The Faculty of Engineering, Memorial University of Newfoundland, St. John's, Canada. (e-mail: licheng@mun.ca).}% <-this % stops a space
\thanks{Manuscript received XX XX, XXXX; revised XX XX, XXXX.}}

\maketitle

\begin{abstract}

In physical layer security (PHY-security), the frequently observed high correlation between the main and wiretap channels can cause a significant loss of secrecy. This paper investigates a slow fading scenario, where a transmitter (Alice) sends a confidential message to a legitimate receiver (Bob) while a passive eavesdropper (Eve) attempts to decode the message from its received signal. It is assumed that Alice is equipped with multiple antennas while Bob and Eve each have a single antenna (i.e., a MISOSE system). In a MISOSE system, high correlation results in nearly collinear main and wiretap channel vectors, which help Eve to see and intercept confidential information. Unfortunately, the signal processing techniques at Alice, such as beamforming and artificial noise (AN), are helpless, especially in the extreme case of completely collinear main and wiretap channel vectors. On this background, we first investigate the achievable secrecy outage probability via beamforming and AN at Alice with the optimal power allocation between the information-bearing signal and AN. Then, an ingenious model, in which a cooperative jamming relay (Relay) is introduced, is proposed to effectively mitigate the adverse effects of high correlation.
Based on the proposed model, the power allocation between the information-bearing signal at Alice and the AN at Relay is also studied to maximize secrecy. Finally, to validate our proposed schemes, numerical simulations are conducted, and the results show that a significant performance gain with respect to secrecy is achieved.

\end{abstract}


% \fi
\IEEEpeerreviewmaketitle

\section{Introduction}
\IEEEPARstart{W}ITH the approach of the so-called big data era, expansive wireless communication networks as critical data contributors \cite{S1-1} have naturally given rise to growing uneasiness about data security. Wireless communication is particularly vulnerable to eavesdropping and impersonation attacks due to the broadcast nature of radio propagation. Traditional security approaches employ symmetric and asymmetric cryptographic algorithms to achieve communication confidentiality and authentication, respectively \cite{S1-2}. Recently, PHY-security techniques have attracted considerable attention as an alternative to the traditional high-complexity cryptography-based secrecy methods.

\par PHY-security has some obvious advantages compared to cryptography-based secrecy methods implemented at the upper layers. For instance, with the rapid advancement of computing technologies, Eve may use infinite computing capabilities to launch brute force attacks or analytical attacks \cite{S1-3}, which can be disastrous for any cryptosystem. By contrast, in PHY-security, the inherent randomness of wireless channels is exploited to guarantee message confidentiality with proper coding and signal processing, which can allow the confidential message to be decoded by only Bob.

\par In the last several decades, researchers have developed a significant number of mathematical theories, technologies, algorithms, and solutions for addressing PHY-security challenges, and the solution varies for each scenario. Almost all the explored signal processing techniques promoting PHY-security aim to increase the signal quality difference at Bob and Eve \cite{S1-4}, and good security performance has been achieved in scenarios where the CSI (channel state information) between the main and eavesdropper channels is independent or weakly correlated. By contrast, scenarios exist where the main and eavesdropper channels are highly correlated, and correlation largely depends on the antenna deployment, the proximity of Bob and Eve, and the scatter around them \cite{S1-5,S1-6,S1-7}. For example, antenna deployment at high altitude in rural or suburban areas generates dominant line-of-sight paths, which results in high correlation between the received signals at the two receivers. Moreover, it is also possible that Eve actively induces correlation, for example, by approaching Bob. Owing to the correlation, the security performance may suffer significant loss. Nevertheless, to the best of our knowledge, few relevant strategies exist and how to strengthen secrecy remains an open issue under adverse conditions. Therefore, we are motivated to develop relevant strategies to strengthen secrecy under conditions of high correlation.

\subsection{Related Work}

The existing literature on PHY-security focus on independent or weakly correlated wiretap channel models;  thus, the involved schemes are not directly applied to the new scenario where the main and wiretap channels are highly correlated. Nevertheless, the related techniques may still be applicable by redesigning the model.

\par In PHY-security, beamforming and precoding techniques at Alice can enhance the signal quality at Bob while limiting the signal strength at Eve. In addition, AN inserted into the transmitted signal can degrade the reception at Eve and consequently further increase the signal quality difference at Bob and Eve. \cite{S1-2} briefly classifies these techniques into four categories, namely, covering beamforming, ZF precoding, convex (CVX)-based precoding, and AN precoding. However, when the main and eavesdropper channels are highly correlated, the signal processing techniques \emph{at Alice} appear to be powerless.

\par Another common technique to improve confidential transmission is relay systems, which can provide additional spatial degrees of freedom through the antennas at the relays. In PHY-security, relays are usually employed to forward data to Bob or to emit AN or jamming signals to disrupt reception at Eve \cite{S1-8,S1-9,S1-10,S1-11,S1-12,S1-13,S1-14,S1-15}. Moreover, relay systems can use full duplex to improve secrecy \cite{S1-16,S1-17,S1-18}. However, these schemes focus on independent or weakly correlated wiretap channel models where channel correlation is not considered.

\par On the other hand, several works address high correlation. \cite{S1-19} propose that secrecy can be enhanced by opportunistically transmitting messages in time slots instead of using excessively large signal power. In particular, confidential transmission occurs when the main channel has better instantaneous channel gain than that of the eavesdropper channel. To maximize the secrecy, power is allocated through a water-filling strategy in the time domain, which states that more power is transmitted in time slots when the channel exhibits high SNR and less power is sent in time slots with poor SNR. Obviously, \cite{S1-19} does not eliminate the fundamental problem caused by high correlation by improving only the usage efficiency of the transmitted power. Additionally, for delay-limited applications, the encoding over multiple channel states adopted in \cite{S1-19} may not be acceptable since it may incur long delays \cite{S1-14}.

\subsection{Scope of Work}

In this paper, we study strategies to improve secrecy in response to pressure from high correlation between the main and eavesdropper channels and propose a scheme using a jamming relay to strengthen the secrecy. This work distinguishes itself from existing literature in following aspects:

\par (1) Traditional beamforming and AN techniques at Alice are exploited to enhance secrecy in a slowly fading Rayleigh environment where the main and wiretap channels are assumed to be highly correlated. Furthermore, the transmitted power allocation between the information-bearing signal and AN at Alice is investigated to further reduce the secrecy outage probability. According to the analysis, high correlation gives rise to significant loss of secrecy, and traditional beamforming and AN at Alice have limited ability to lower the adverse effects of high correlation.

\par (2)  To overcome the secrecy performance degradation due to high correlation, a cooperative jamming relay is introduced into the system to create new conditions for confidential transmission. By employing the cooperative jamming relay to emit AN to disrupt reception at Eve, the difference at Bob and Eve is increased, consequently improving the secrecy performance. This strategy depends on only instantaneous characteristic channel gain, which does not incur long delays.

\par (3)  The joint power allocation between the information-bearing signal at Alice and AN at Relay is presented to further enhance the confidentiality of the system. For this purpose, we propose and discuss the optimal power allocation ratio between Alice and Relay. Based on the results, the secrecy outage probability in the proposed scheme with a cooperative jamming relay is computed and simulated. Compared to the traditional scheme, low secrecy outage probability is ensured.

\subsection{Outline of the Paper}

\par The rest of this paper is organized as follows: section II presents the system model with high correlation between the main and wiretap channels; in section III, traditional beamforming and AN at Alice are exploited to enhance secrecy under high correlation, and the transmitted power allocation between the information-bearing signal and AN at Alice is investigated to lower the secrecy outage probability;
in section IV, we propose a scheme that introduces a cooperative jamming relay into the system to create new conditions for confidential transmission, which is followed by joint power allocation between the information-bearing signal at Alice and AN at Relay; section V presents simulations whose numerical results
demonstrate that the proposed schemes provide significant performance gain when the main and wiretap channels are highly correlated; section VI provides the conclusions of this work.

\par Throughout this paper, the following notation will be used: Boldface upper and lower cases denote matrices and vectors, respectively. ${[ \cdot ]^H}$ denotes the conjugate transpose operation. The notation $E[ \cdot ]$ denotes the mathematical expectation. $\left|  \cdot  \right|$ denotes the norm of a vector.




\section{System Model}

As illustrated in Fig. 1, we consider the wireless scenario where Alice sends a confidential message to Bob while the transmission is overheard by only one passive Eve. It is assumed that Alice is equipped with \emph{M} antennas while Bob and Eve each have a single antenna. Since Eve passively receives the transmitted signal from Alice, Alice can only obtain the CSI of the Alice-Bob link when the CSI of the Alice-Eve link is not known. However, it is reasonable to assume that the statistical CSI of the Alice-Eve link is available to Alice since these statistics can be obtained from prior measurements of the environment. Here, we directly assume that the Alice-Bob and Alice-Eve links are slowly fading Rayleigh channels. Note that the secrecy of communications is not dependent on the secrecy of channel gain; therefore, all the channel gains can be published.

\begin{figure}
\centering
\includegraphics[width=3.0in]{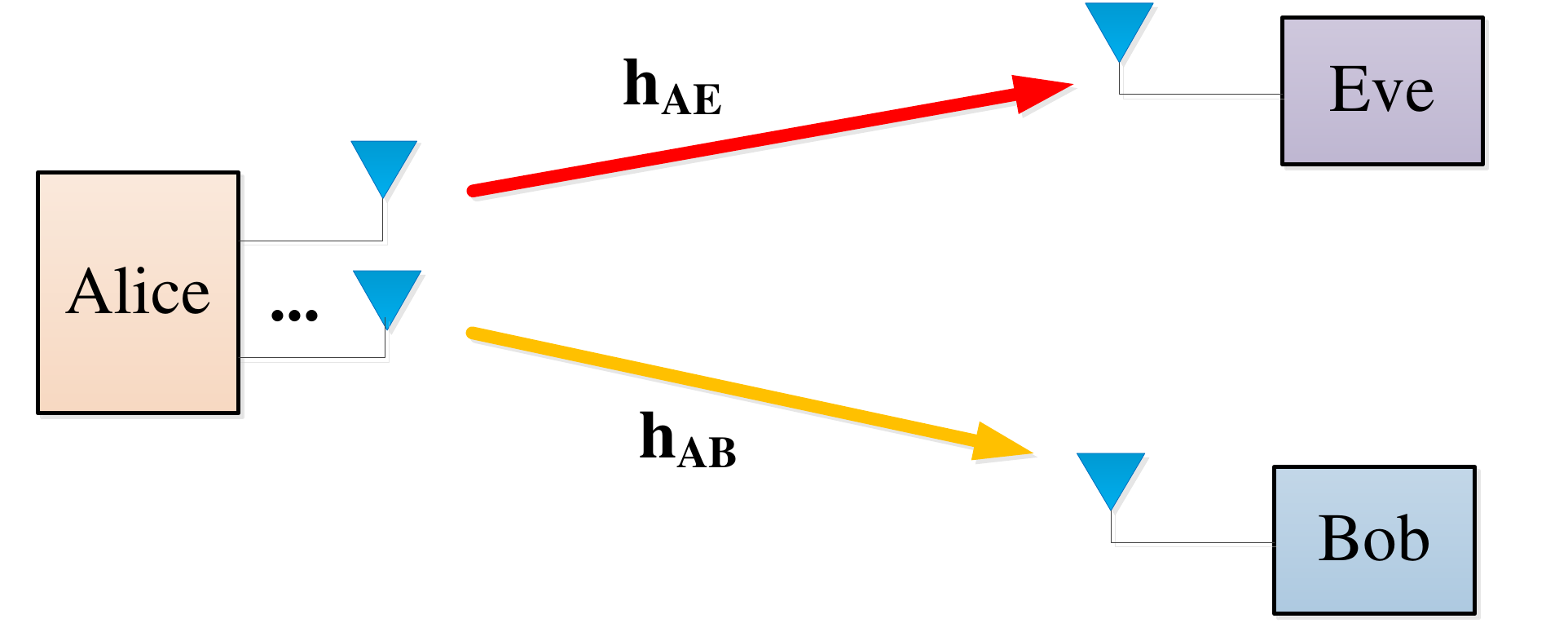}
\caption{System model.}
\label{Fig1}
\end{figure}

\par When Alice transmits a data stream bearing useful information, the signals received at Bob and Eve are written, respectively, as

\begin{equation}\label{Equation8}
{y_B} = {\bf{h}}_{AB}^H{\bf{x}} + {{n}_B}
\end{equation}

\noindent and

\begin{equation}\label{Equation8}
{y_E} = {\bf{h}}_{AE}^H{\bf{x}} + {{n}_E},
\end{equation}

\noindent where ${\bf{x}} \in {{\bf{C}}^{M \times 1}}$ is the transmitted signal vector from Alice and $E\{ |{\bf{x}}{|^2}\}  \le P$ represents the total power of the transmitted signal from Alice. ${{\bf{h}}_{AB}}$ and ${{\bf{h}}_{AE}}$ denote the channel gain vectors from Alice to Bob and from Alice to Eve, respectively, with ${{\bf{h}}_{AB}}\sim \mathcal{CN}(0,{\sigma _{AB}}{\bf{I}})$ and ${{\bf{h}}_{AE}}\sim {\cal C}{\cal N}(0,{\sigma _{AE}}{\bf{I}})$ assumed. ${{n}_B}$ and ${{n}_E}$ denote the additive noise at Bob and Eve, respectively, with ${{n}_B}\sim \mathcal{CN}(0,1)$ and ${{n}_E}\sim \mathcal{CN}(0,1)$ assumed. It is worth noting that the path loss related to the distance is modeled as position-dependent channel gain variance, whereas the power of the additive noise is normalized for simplification.

\par In a certain regional scope of a wireless environment, wireless channels between different users may be associated with each other, the extent of which often depends on the distance between users and the scatter around them. In our model,to intercept more confidential information, Eve may actively approach Bob to increase the correlation between ${{\bf{h}}_{AB}}$ and ${{\bf{h}}_{AE}}$. The reason is that the distance from Alice to Bob and Eve is far larger than that between Bob and Eve, which may result in a similar signal path. Due to the high correlation, the difference between the main and wiretap channels is very small. We use the correlation coefficient between ${{\bf{h}}_{AB}}$ and ${{\bf{h}}_{AE}}$ (denoted by $\rho$) to measure the degree of correlation between the main and wiretap channels. In particular, $\rho$ can be expressed as follows,

\begin{equation}\label{Equation8}
\rho  = \frac{{{{\bf{h}}_{AB}} \cdot {{\bf{h}}_{AE}}}}{{\left| {{{\bf{h}}_{AB}}} \right|\left| {{{\bf{h}}_{AE}}} \right|}},
\end{equation}

\noindent where $0 \leq \rho \leq 1$, with $\rho$ = 0 indicating that ${{\bf{h}}_{AB}}$ and ${{\bf{h}}_{AE}}$ are completely uncorrelated and $\rho$ = 1 representing full correlation in a time slot. On the other hand, since the entries of ${{\bf{h}}_{AB}}$ and ${{\bf{h}}_{AE}}$ are independent and identically distributed (i.i.d.) and drawn from a complex Gaussian distribution with zero mean, ${\left| {{{\bf{h}}_{AB}}} \right|^2}$  and ${\left| {{{\bf{h}}_{AE}}} \right|^2}$  both follow Gamma distributions with parameters $(2M, {\sigma _{AB}}/\sqrt 2)$ and $(2M, {\sigma _{AE}}/\sqrt 2)$, respectively.

\par Since wireless environments often change very slowly, it is reasonable to assume that the correlation between ${{\bf{h}}_{AB}}$ and ${{\bf{h}}_{AE}}$ can be estimated through multiple prior measurements of the environment and feedback coming from other users. Here, we assume that $\rho$ can be attained by a series of processes, whereas the values of ${\left| {{{\bf{h}}_{AB}}} \right|}$  and ${\left| {{{\bf{h}}_{AE}}} \right|}$ are unknown and independent of each other. It is worth emphasizing that $\rho$ represents the relationship between two ``spatial'' vectors, in contrast to the usual relationship between ``temporal'' signal vectors.

\section{secrecy performance under high correlation}

\par In a MISOSE system, beamforming at Alice is employed as a common PHY-security technique to enhance the signal quality at Bob while limiting the signal strength at Eve. By contrast, AN techniques can be used to degrade the reception at Eve to increase the signal quality difference at Bob and Eve. In general, when Eve's CSI is unknown, the information-bearing signal can only be directed towards Bob, and AN can effectively facilitate secrecy \cite{S1-4}. In this section, we discuss how these techniques affect the secrecy performance under the system model described above.

\par Under the system model described above, Alice simultaneously transmits an information-bearing signal and AN
to improve the secrecy performance. When AN is added to the null space of ${{\bf{h}}_{AB}}$, the transmitted signal vector \textbf{x} can be modified as follows,

\begin{equation}\label{Equation8}
\begin{array}{*{20}{l}}
{\bf{x}}{\rm{ = }}\sqrt {\phi P} {\bf{f}}u + \sqrt {\dfrac{{1 - \phi }}{{M - 1}}P} {{\bf{a}}_A},
\end{array}��
\end{equation}

\noindent where $u\sim \mathcal{CN}(0,1)$ corresponds to symbols in a Gaussian codebook; $\bf{f}$ represents a beamformer; the transmitted signal is directed at Bob with $\bf{f} = {{\bf{h}}_{AB}}/\left| {{{\bf{h}}_{AB}}} \right|$ (this beamformer design maximizes the average secrecy capacity when Eve's CSI is unknown to Alice \cite{S3-1}); ${{\bf{a}}_A}$ denotes AN embedded into the transmitted signal with ${{\bf{a}}_A}\sim {\cal C}{\cal N}(0,\bf{I})$; the ratios of power allocated to the information-bearing signal and the AN are denoted as $\phi$ and $1-\phi$, respectively; and the remaining symbols have the same or similar meaning as before.

\par Since Eve's CSI is unknown, the AN is chosen within the null space of ${{\bf{h}}_{AB}}$ to prevent AN leakage into main channel, such that ${\bf{h}}_{AB}^H{{\bf{a}}_A} = 0$. By contrast, some components of the AN may lie in the range space of the Alice-Eve link, which disrupts the reception at Eve with high probability.  Therefore, the received signals at Bob and Eve are rewritten, respectively, as

\begin{equation}\label{Equation8}
{y_B} = \sqrt {\phi P} {\bf{h}}_{AB}^H{\bf{f}}u + {n_B},
\end{equation}

\noindent and

\begin{equation}\label{Equation8}
{y_E} = \sqrt {\phi P} {\bf{h}}_{AE}^H{\bf{f}}u + \sqrt {\dfrac{{1 - \phi }}{{M - 1}}P}{\bf{h}}_{AE}^H{{\bf{a}}_A} + {n_E}.
\end{equation}

\par Since ${{\bf{h}}_{AE}}$ is not known to Alice, a fixed ${{\bf{a}}_A}$ may result in a case where ${\left| {{\bf{h}}_{AE}^H{{\bf{a}}_A}} \right|^2}$ is small; thereby, the power of AN seen by Eve is also small. To reduce or avoid this possibility, it may be better to randomly choose ${{\bf{a}}_A}$ as a complex Gaussian vector in the null space of ${{\bf{h}}_{AB}}$ \cite{S3-1}. In particular, ${{\bf{a}}_A}$ can be generated through ${{\bf{a}}_A} = {{\bf{Z}}_{AB}}{{\bf{v}}_A}$, where ${{\bf{Z}}_{AB}}$ is an orthonormal basis for the null space of ${{\bf{h}}_{AB}}$ with ${\bf{Z}}_{AB}^H{{\bf{Z}}_{AB}} = {\bf{I}}$, and the \emph{M}-1 entries of ${{\bf{v}}_A}$ are i.i.d. values drawn from a complex Gaussian distribution with zero mean and unit variance. Obviously, the power of the AN signal is equally distributed in the null space of ${{\bf{h}}_{AB}}$.

\subsection{Secrecy Performance}
\par The SNRs at Bob and Eve are rewritten, respectively, as

\begin{equation}\label{Equation8}
SN{R_B} = \phi P{\left| {{{\bf{h}}_{AB}}} \right|^2},
\end{equation}

\noindent and

\begin{equation}\label{Equation8}
SN{R_E} = \dfrac{{\phi P{{\left| {{{\bf{h}}_{AE}}^H{\bf{f}}} \right|}^2}}}{{1 + \dfrac{{1 - \phi }}{{M - 1}}P|{{\bf{h}}_{AE}}^H{{\bf{a}}_A}{|^2}}}.
\end{equation}

\noindent Since the power of the AN signal is equally distributed in the null space of ${{\bf{h}}_{AB}}$, by
considering the correlation coefficient $\rho$, Eq. 8 can be rewritten as

\begin{equation}\label{Equation8}
SN{R_E} = \dfrac{{\phi {\rho ^2}P{{\left| {{{\bf{h}}_{AE}}} \right|}^2}}}{{1 + \dfrac{{1 - \phi }}{{M - 1}}(1 - {\rho ^2})P|{{\bf{h}}_{AE}}{|^2}}}.
\end{equation}

\par Thus, the capacity of the main channel is given by

\begin{equation}\label{Equation8}
\begin{aligned}
{C_m} & = \log (1 + SN{R_B})\\
      & = \log (1 + \phi P{\left| {{{\bf{h}}_{AB}}} \right|^2}),
\end{aligned}
\end{equation}

\noindent and that of the wiretap  channel is

\begin{equation}\label{Equation8}
\begin{aligned}
{C_w} & = \log (1 + SN{R_E})\\
& = \log (1 + \dfrac{{\phi \rho^2 P{{\left| {{{\bf{h}}_{AE}}} \right|}^2}}}{{1 + \dfrac{{1 - \phi }}{{M - 1}}(1 - {\rho ^2})P|{{\bf{h}}_{AE}}{|^2}}}).
\end{aligned}
\end{equation}

\noindent Based on Eq. 10 and Eq. 11, the secrecy capacity over a block consisting of a large number of symbols can be obtained as

\begin{equation}{\label {eq8}}
\begin{aligned}
C_{s} = &[C_{m}-C_{w}]^{+}\\
 =& \max \{C_{m}-C_{w},0\}.
\end{aligned}
\end{equation}

\par In our proposed model, the secrecy outage probability and outage capacity are considered when ideal interleaving is impossible due to strict delay restrictions, and the channel capacity cannot be expressed as the average of the capacities for all possible channel realizations (ergodic secrecy capacities). We characterize the outage probability as follows,

\begin{equation}\label{Equation8}
{P_{out}}({R_s}) = P\{{C_s} < {R_s}\},
\end{equation}

\noindent i.e. the probability that the instantaneous secrecy capacity is less than the target secrecy rate $R_s$. In addition, $R_s$ is the highest secrecy transmission rate that maintains the secrecy outage probability under the given ${P_{out}}$, i.e., the outage capacity subject to ${P_{out}}$. According to Eqs. 10 - 12, the outage probability can be rewritten as

\begin{equation}\label{Equation8}
{P_{out}}({R_s}) = P\{SN{R_B} - {2^{{R_s}}}SN{R_E} - {2^{{R_s}}} + 1  < 0\},
\end{equation}

\noindent where $SN{R_B}$ and $SN{R_E}$ are given in Eq. 7 and Eq. 9, respectively. Further, given a target secrecy rate $R_s$, the secrecy outage probability is obtained via Monte Carlo simulations.

\subsection{Optimal Power Allocation}

\par Based on the derivation of ${P_{out}}$, it is not difficult to see that the ratios of the power allocation
$\phi$ and ${{\bf{h}}_{AB}}$ are key factors in the secrecy outage probability and outage capacity. In the following, we optimize the transmitted power allocation according to the instantaneous realization of ${{\bf{h}}_{AB}}$ and the statistical distribution of ${{\bf{h}}_{AE}}$, aiming to minimize the secrecy outage probability under a target secrecy rate $R_s$. Mathematically, the problem can be expressed as

\begin{equation}{\label {eq6}}
\begin{aligned}
&\max  \quad\quad \{ {{\rm{C}}_m} - E[{C_w}]\} \\
&s.t.\quad\quad\ E\{ |{\bf{x}}{|^2}\}  \le P,\\
&\quad\quad\quad\quad 0\leq \phi \leq1,
\end{aligned}
\end{equation}

\noindent where $E[{C_w}]$ is the expectation of the wiretap capacity over ${{\bf{h}}_{AE}}$, i.e.,

\begin{equation}{\label {eq6}}
\begin{aligned}
E[{C_w}] & = {E_{{{\bf{h}}_{AE}}}}\{ \log (1 + \dfrac{{\phi {\rho ^2}P{{\left| {{{\bf{h}}_{AE}}} \right|}^2}}}{{1 + \dfrac{{1 - \phi }}{{M - 1}}(1 - {\rho ^2})P|{{\bf{h}}_{AE}}{|^2}}})\} \\
& = {E_{{{\bf{h}}_{AE}}}}\{ \log (\dfrac{{1 + [\dfrac{{1 - \phi }}{{M - 1}}(1 - {\rho ^2}) + \phi {\rho ^2}]P{{\left| {{{\bf{h}}_{AE}}} \right|}^2}}}{{1 + \dfrac{{1 - \phi }}{{M - 1}}(1 - {\rho ^2})P|{{\bf{h}}_{AE}}{|^2}}})\}\\
& \buildrel \Delta \over = {C_{w1}} - {C_{w2}},
\end{aligned}
\end{equation}

\noindent where

\begin{equation}{\label {eq6}}
\begin{aligned}
{C_{w1}} = {E_{{{\bf{h}}_{AE}}}}\{ \log (1 + [\frac{{1 - \phi }}{{M - 1}}(1 - {\rho ^2}) + \phi {\rho ^2}]P{\left| {{{\bf{h}}_{AE}}} \right|^2})\} ,
\end{aligned}
\end{equation}

\noindent and

\begin{equation}{\label {eq6}}
\begin{aligned}
{C_{w2}} = {E_{{{\bf{h}}_{AE}}}}\{ \log (1 + \frac{{1 - \phi }}{{M - 1}}(1 - {\rho ^2})P{\left| {{{\bf{h}}_{AE}}} \right|^2})\} ,
\end{aligned}
\end{equation}

\noindent where ${\left| {{{\bf{h}}_{AE}}} \right|^2}$ follows a Gamma distribution whose probability density function (PDF) satisfies

\begin{equation}{\label {eq6}}
\begin{aligned}f(x) = \dfrac{1}{{{\sigma _{AE}}^2}}\frac{1}{{\Gamma (M)}}{(\frac{x}{{{\sigma _{AE}}^2}})^{M - 1}}{e^{ - \dfrac{x}{{{\sigma _{AE}}^2}}}}.
\end{aligned}
\end{equation}

\par Based on Eqs. 19 - 21, ${C_{w1}}$ and ${C_{w2}}$ are given, respectively, by

\begin{equation}{\label {eq6}}
\begin{aligned}
{C_{w1}} &=\dfrac{1}{{\ln 2}}\exp (\dfrac{1}{{[\dfrac{{1 - \phi }}{{M - 1}}(1 - {\rho ^2}) + \phi {\rho ^2}]{\sigma _{AE}}^2P}}) \cdot\\
& \quad \quad \quad \sum\limits_{n = 1}^M {{E_n}} (\dfrac{1}{{[\dfrac{{1 - \phi }}{{M - 1}}(1 - {\rho ^2}) + \phi {\rho ^2}]{\sigma _{AE}}^2P}})
\end{aligned}
\end{equation}

\noindent and

\begin{equation}{\label {eq6}}
\begin{aligned}
{C_{w2}} &= \dfrac{1}{{\ln 2}}\exp (\dfrac{1}{{\dfrac{{1 - \phi }}{{M - 1}}(1 - {\rho ^2}){\sigma _{AE}}^2P}})\cdot\\
& \quad \quad \quad  \sum\limits_{n = 1}^M {{E_n}} (\dfrac{1}{{\dfrac{{1 - \phi }}{{M - 1}}(1 - {\rho ^2}){\sigma _{AE}}^2P}}).
\end{aligned}
\end{equation}

\par It is worth noting that Eq. 20 and Eq. 21 are obtained using the formula given in \cite{S3-3}, i.e.,

\begin{equation}{\label {eq6}}
\begin{aligned}
\int_0^\infty  {log(1 + \alpha x)} \dfrac{{{x^{N - 1}}}}{{\Gamma (N){e^x}}}dx = \dfrac{1}{\ln2} \exp (\frac{1}{\alpha })\sum\limits_{n = 1}^N {{E_n}(\frac{1}{\alpha })},
\end{aligned}
\end{equation}

\noindent where %EDITOR: Please consider whether a variable is missing here.
 is the exponential integral of the order defined by

\begin{equation}{\label {eq6}}
\begin{aligned}
{E_n}(\frac{1}{\alpha } ) = \int_1^{ + \infty } {{t^{ - n}}{e^{ - \frac{t}{\alpha }}}dt, \quad    {\rm{      }}\frac{1}{\alpha }}  \ge 0.
\end{aligned}
\end{equation}

\par According to Eqs. 20 and 21, the expectation of the wiretap capacity over ${{\bf{h}}_{AE}}$, i.e., $E[{C_w}]$, can be computed for any value of the ratio of power allocation $\phi$.

\begin{figure}
\centering
\includegraphics[width=3.3in]{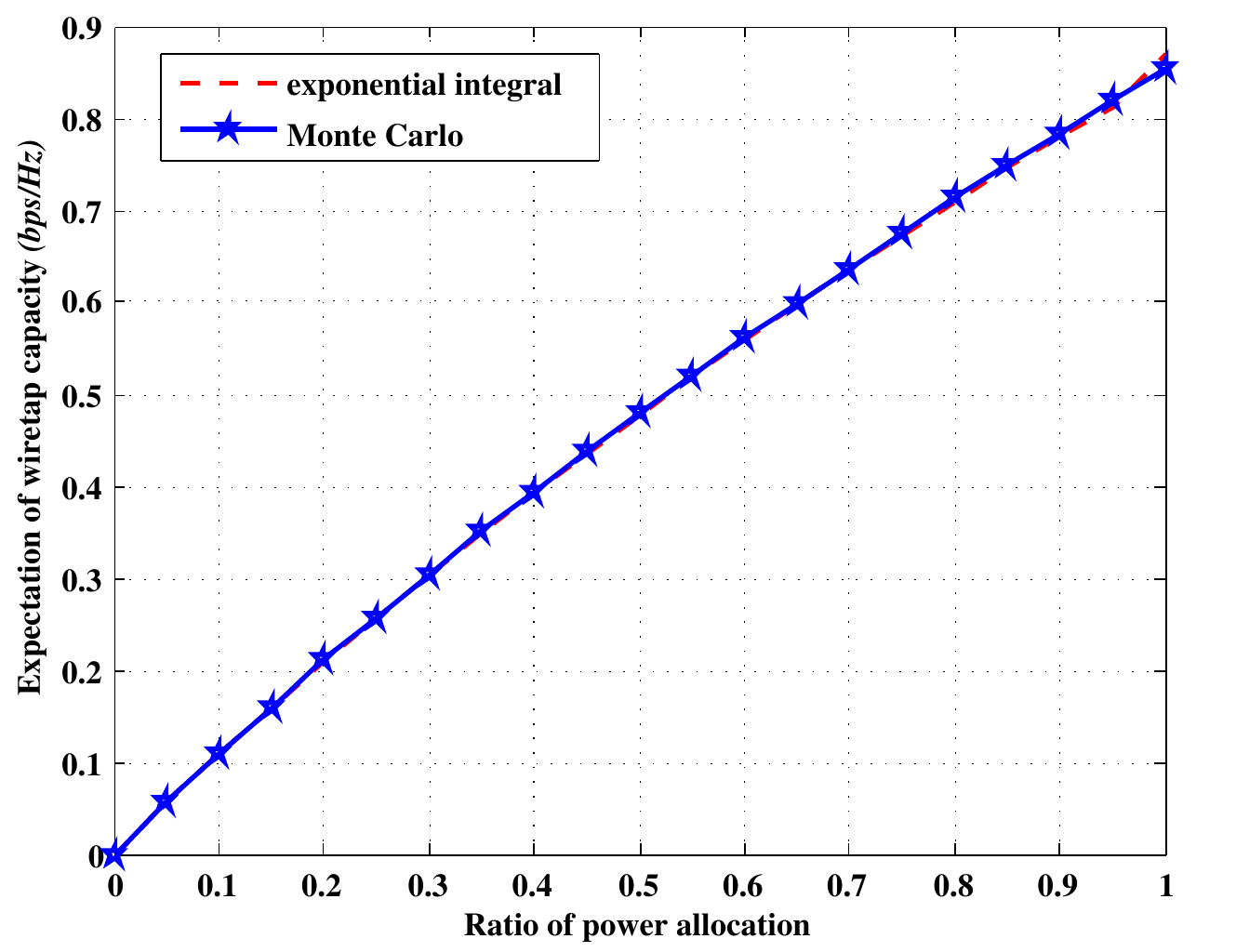}
\caption{The expectation of the wiretap capacity over ${{\bf{h}}_{AE}}$ versus the ratio of power allocation.}
\label{Fig}
\end{figure}

\par Fig. 2 shows that the result curve obtained through Monte Carlo simulation perfectly agrees with that obtained via the aforementioned method using the  integral of exponential order, confirming the accuracy of the two methods. It is worth noting that the method using the integral of exponential order has lower computational complexity than that of the Monte Carlo simulation. Therefore, the former is appropriate for the case where derivation can be achieved, whereas the latter can be applied in the opposite case of high computational complexity.

\par Based on these results, the maximum instantaneous secrecy capacity can be achieved by selecting the optimal $\phi$. Then, the secrecy outage probability $P_{out}$ can be obtained via Monte Carlo simulation.

%The procedure computing $\phi$ and $P_{out}$ is summarized in \textbf{Algorithm 1}. The computation complexity of this algorithm is \textcolor[rgb]{1.00,0.00,0.00}{low (O(N2))}, making it quick to execute.

\section{Improving Secrecy using a jamming relay}

In many wireless situations, the difference between ${{\bf{h}}_{AB}}$ and ${{\bf{h}}_{AE}}$ is exploited to achieve a positive secrecy rate, even if the superiority of the main channel is not guaranteed. In this process, signal processing techniques, including precoding/beamforming and AN, are often employed. Precoding/beamforming techniques can enhance the signal quality at Bob while limiting the signal strength at Eve. AN techniques can degrade the reception at Eve, thereby increasing the signal quality difference at Bob and Eve. However, if the difference between ${{\bf{h}}_{AB}}$ and ${{\bf{h}}_{AE}}$ is small, the loss of secrecy capacity due to correlation is increased. In this case, the excessively large signal power does not help to improve the secrecy capacity \cite{S1-19}, and the usual schemes appear to be powerless. To overcome this disadvantage, we propose an ingeniously designed scheme where a cooperative jamming relay is introduced into the system to create new conditions for confidential transmission, as shown in Fig 3.

\begin{figure}
\centering
\includegraphics[width=3.0in]{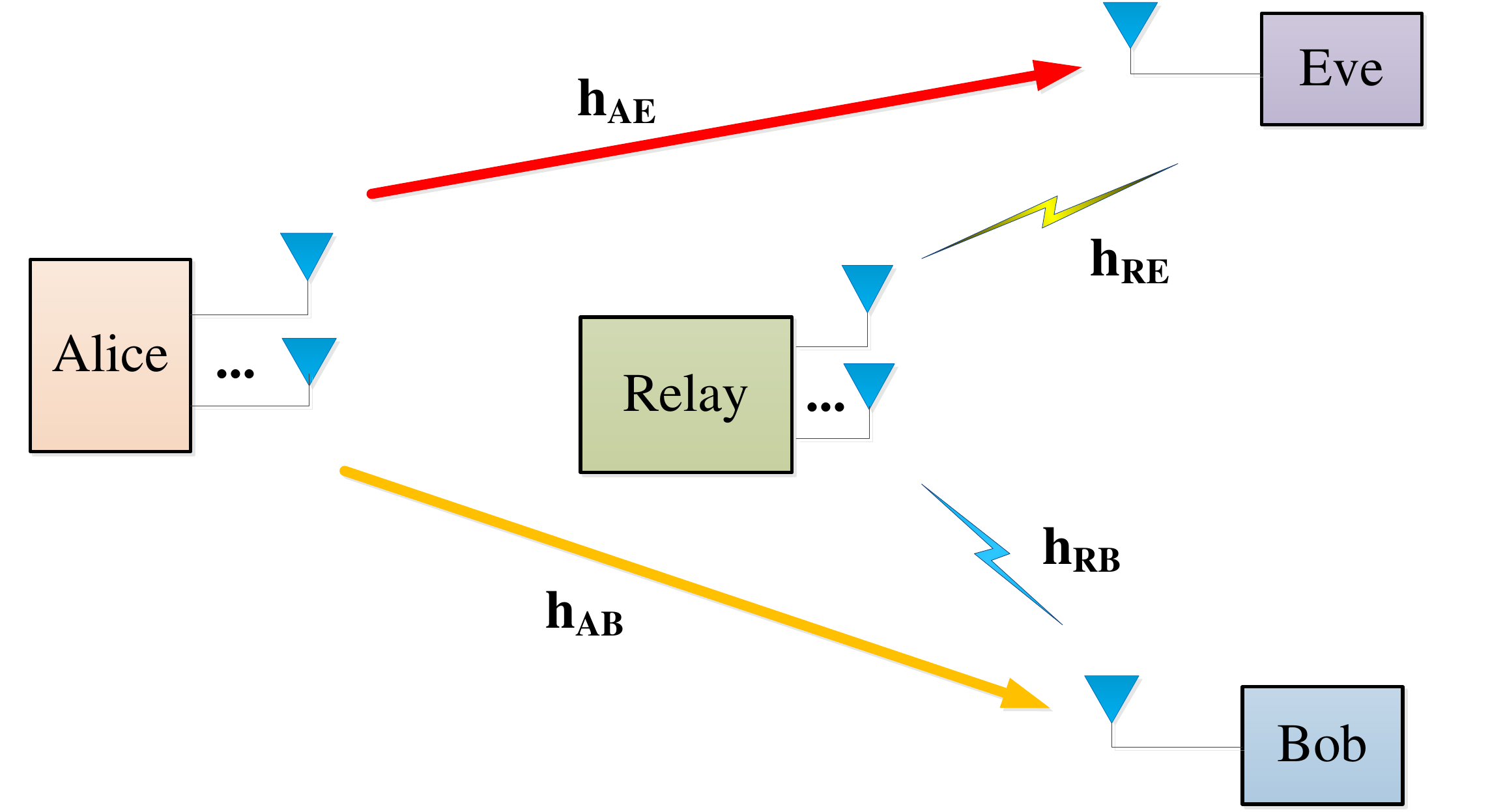}
\caption{Improving secrecy by introducing a jamming relay.}
\label{Fig2}
\end{figure}

\par When Relay transmits signals that are independent of the intended message, these signals create interference for both Bob and Eve, limiting both of their decoding capabilities and reducing their reliable decoding rates. However, the net effect of this jamming may be an increase in the difference of the rates and hence an increase in the achievable secrecy rate of the Alice-Bob link. In other words, if an independent transmission jams Bob and Eve simultaneously, Bob may experience a net gain.

\par In our proposed scheme, Relay is equipped with \emph{N} antennas. In general, the Relay-Bob and Relay-Eve links (denoted by ${{\bf{h}}_{RB}}$ and ${{\bf{h}}_{RE}}$, respectively) are uncorrelated due to a reasonable deployment of Relay. In many wireless environments, it is reasonable to assume that the Relay-Bob and Relay-Eve links are slowly fading Rayleigh channels. To enhance the confidentiality, Relay emits AN to disrupt the reception at Eve. Since ${{\bf{h}}_{RB}}$ and ${{\bf{h}}_{RE}}$ are uncorrelated, signal processing techniques, including precoding/beamforming and AN, are applicable to this new scenario. Although Eve could attempt to mitigate the interference signal from Relay using a directional antenna, ${{\bf{h}}_{AE}}$ can be significantly changed accordingly to amplify the difference between ${{\bf{h}}_{AB}}$  and ${{\bf{h}}_{AE}}$. In this case, the difficulty caused by the high correlation between the received signals at Bob and Eve disappears; therefore, the case is not considered.

\par We assume that Relay has the CSI of the Relay-Bob link and that the statistical CSI of the Relay-Eve link is available to Relay. Similar to the use of AN at Alice in Section III, Relay transmits AN in the null space of the Relay-Bob link with ${{\bf{h}}_{RB}}^H{{\bf{a}}_R} = 0$, and some components of AN may lie in the range space of the Relay-Eve link. The received signals at Bob and Eve, jointly considering the signal containing a confidential message from Alice and AN from Relay, are written, respectively, as

\begin{equation}\label{Equation8}
{y_B} = \sqrt {\lambda P} {{\bf{h}}_{AB}}^H{\bf{f}}u  + {n_B},
\end{equation}

\noindent and

\begin{equation}\label{Equation8}
{y_E} = \sqrt {\lambda P} {{\bf{h}}_{AE}}^H{\bf{f}}u + \sqrt {\frac{{1 - \lambda }}{{N - 1}}P} {{\bf{h}}_{RE}}^H{{\bf{a}}_R} + {n_E},
\end{equation}

\noindent where ${{\bf{a}}_R}$ denotes the AN vector in the null space of ${{\bf{h}}_{RB}}$ with ${{\bf{a}}_R}\sim {\cal C}{\cal N}(0,N - 1)$; $\lambda$ represents the fraction of power allocated to Alice; and the rest of the symbols have the same or similar meaning as before. Note that in the model, Alice only transmits the information-bearing signal without AN, i.e., ${\bf{x}} = {\bf{f}}u$, because 1) the distance between Relay and Eve is far less than that between Alice and Eve, which results in lower power consumption over the travel path; 2) the correlation between ${{\bf{h}}_{RB}}$ and ${{\bf{h}}_{RE}}$ is small, which guarantees high jamming efficiency; and 3) the equipment complexity at Alice is reduced significantly.

\subsection{Secrecy Performance}

\par The SNRs at Bob and Eve are rewritten, respectively, as

\begin{equation}\label{Equation8}
SN{R_B} = \lambda P{\left| {{{\bf{h}}_{AB}}} \right|^2},
\end{equation}

\noindent and

\begin{equation}\label{Equation8}
\begin{aligned}
SN{R_E} & = \dfrac{{\lambda \rho^2 P{{\left| {{{\bf{h}}_{AE}}} \right|}^2}}}{{1 + \dfrac{{1 - \lambda }}{{N - 1}}P|{{\bf{h}}_{RE}} \cdot {{\bf{a}}_R}|^2}}\\
& = \frac{{\lambda {\rho ^2}P{{\left| {{{\bf{h}}_{AE}}} \right|}^2}}}{{1 + \dfrac{{1 - \lambda }}{{N - 1}}(1 - \rho _R^2)P|{{\bf{h}}_{RE}}{|^2}}},
\end{aligned}
\end{equation}

\noindent where

\begin{equation}\label{Equation8}
{\rho _R} = \dfrac{{{{\bf{h}}_{RE}} \cdot {{\bf{h}}_{RB}}}}{{\left| {{{\bf{h}}_{RE}}} \right|\left| {{{\bf{h}}_{RB}}} \right|}}.
\end{equation}

\par Thus, the capacity of the main channel is given by

\begin{equation}\label{Equation8}
\begin{aligned}
{C_m} = \log (1 + \lambda P{\left| {{{\bf{h}}_{AB}}} \right|^2}),
\end{aligned}
\end{equation}

\noindent and that of the wiretap  channel is

\begin{equation}\label{Equation8}
\begin{aligned}
{C_w} = \log (1 + \frac{{\lambda {\rho ^2}P{{\left| {{{\bf{h}}_{AE}}} \right|}^2}}}{{1 + \dfrac{{1 - \lambda }}{{N - 1}}(1 - \rho _R^2)P|{{\bf{h}}_{RE}}{|^2}}}).
\end{aligned}
\end{equation}

\par On this basis, we can obtain the secrecy capacity over a block consisting of a large number of symbols and the outage probability according to Eq. 12 and Eq. 14, respectively.

\subsection{Joint Power Allocation of Alice and Relay }

\par When the level of the correlation between the main and wiretap channels is very high, the secrecy performance of the system suffers significant loss. In view of this situation, we present a scheme to introduce a cooperative jamming relay. Although the relay generates a new dimension to overcome the secrecy performance degradation caused by high correlation, it is obvious that better power allocation between Alice and Relay plays a essential role in guaranteeing a lower secrecy outage probability. Thus, in this section, we focus on the joint power allocation of Alice and Relay.

\par Similar to section III.\emph{B}, in this scheme, the problem of minimizing the secrecy outage probability under a target secrecy rate $R_s$ can be expressed as

\begin{equation}{\label {eq6}}
\begin{aligned}
&\max  \quad\quad \{ {{\rm{C}}_m} - E[{C_w}]\} \\
&s.t.\quad\quad\ E\{ |{\bf{x}}{|^2}\}  \le P,\\
&\quad\quad\quad\quad 0\leq \lambda \leq1,
\end{aligned}
\end{equation}

\noindent where $E[{C_w}]$ is the expectation of the wiretap capacity over ${{\bf{h}}_{AE}}$ and ${{\bf{h}}_{RE}}$, i.e.,

\begin{equation}{\label {eq6}}
\begin{aligned}
E[{C_w}] &= {E_{{{\bf{h}}_{AE}},{{\bf{h}}_{RE}}}}\{ \log (1 + \dfrac{{\lambda {\rho ^2}P{{\left| {{{\bf{h}}_{AE}}} \right|}^2}}}{{1 + \dfrac{{1 - \lambda }}{{N - 1}}(1 - \rho _R^2)P|{{\bf{h}}_{RE}}{|^2}}}).
\end{aligned}
\end{equation}

\par The expectation of the wiretap capacity in Eq. 32, i.e., $E[{C_w}]$, can be obtained via Monte Carlo simulations for any power allocation ratio $\phi$. Based on these results, the maximum instantaneous secrecy capacity can be achieved by choosing the optimal $\phi$. Then, the secrecy outage probability $P_{out}$ can be obtained using Monte Carlo simulations.

\section{NUMERICAL RESULTS}

In this section, computer simulations are performed to evaluate the performance of our proposed scheme of introducing a cooperative jamming relay. We start by investigating how the square of the correlation coefficient between the main and wiretap channels affects the secrecy outage probability; the traditional scheme employing beamforming and AN at Alice is presented for comparison. Then, we present the secrecy outage probability with respect to the total transmitted power. Lastly, the secrecy outage probability versus the target secrecy rate $R_s$ or outage capacity is given.

\subsection{$P_{out}$ versus $\rho^2$}

Here, we discuss the relationship between the secrecy outage probability and the correlation coefficient. Some of the parameters in the simulations are set as follows: the total transmitted power \emph{P} is 3 dBW; the target secrecy rate $R_s$ is 0 bps/Hz since we do not focus on a specified communication in practice; the additive noise at Bob and Eve is assumed to be i.i.d. drawn from a complex Gaussian distribution with zero mean and unit variance; the variances of ${{\bf{h}}_{AB}}$ and ${{\bf{h}}_{AE}}$, which are related to the path loss depending on distance, are both ${\sigma _{AB}^2} = {\sigma _{AE}^2} = 1/2$ because Bob and Eve are very close. It is worth noting that owing to the normalized noise power and $P = P/\sigma_n$, the total transmitted power can be measured in "dB".

\begin{figure}
\centering
\includegraphics[width=3.6in]{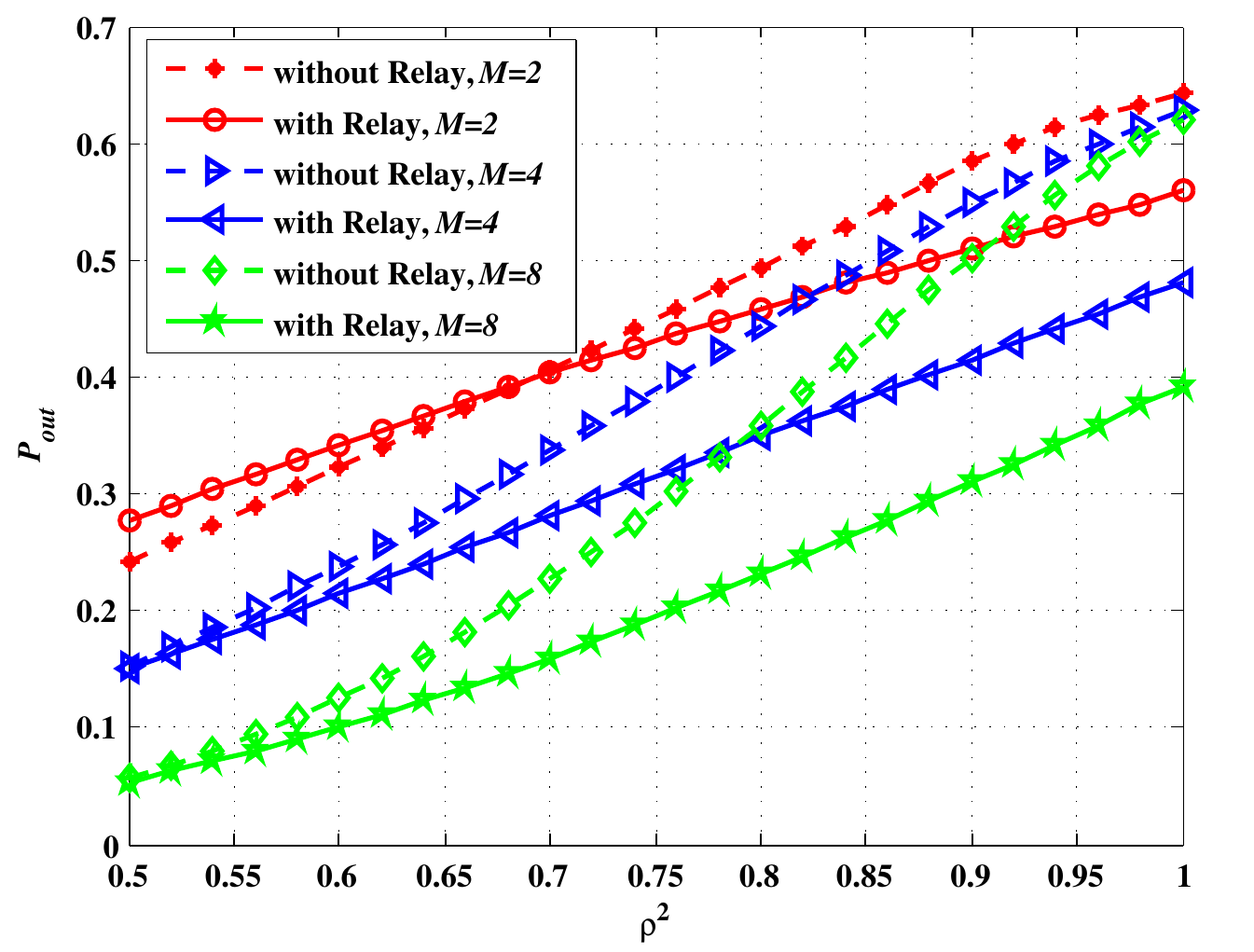}
\caption{The relationship between the secrecy outage probability $P_{out}$ and the square of the correlation coefficient between the main and wiretap channels $\rho^2$.}
\label{Fig3}
\end{figure}

\par Fig. 4 plots the relationship between the secrecy outage probability $P_{out}$ and the square of the correlation coefficient between the main and wiretap channels $\rho^2$. Relay is generally deployed closer to Bob than to Alice; therefore, the path loss between Relay and Bob is lower than that from Alice to Bob and can be modeled as ${\sigma _{RB}^2} > {\sigma _{AB}^2}$. However, to exclude the benefit of the shorter distance between Relay and Bob and to focus on the effects of introducing a cooperative jamming relay, we intentionally set ${\sigma _{RB}^2} = {\sigma _{AB}^2}$ to ensure an equivalent path loss. In addition, we set ${\sigma _{RB}^2} = {\sigma _{RE}^2}$ because Bob and Eve are very close. Fig. 4 presents three cases, where the number of antennas at Alice \emph{M} = 2, 4 and 8 and the number of antennas at Relay is \emph{N} = 2 . Fig. 4 clearly shows that the secrecy outage probability depends largely on $\rho^2$. When $\rho^2$ is very large, the secrecy outage probability is also large. In the extreme case where ${{\bf{h}}_{AB}}$ and ${{\bf{h}}_{AE}}$ are completely collinear, i.e., $\rho^2 = 1$, the secrecy outage probability is approximately 0.63 and only depends on $\left| {{{\bf{h}}_{AB}}} \right|$ and $\left| {{{\bf{h}}_{AE}}} \right|$. The secrecy outage probability is not 0.5 because the optimal power allocation ratio between Alice and Relay is selected based on the instantaneous secrecy capacity rather than the secrecy outage probability. Our proposed scheme using a cooperative jamming relay has a lower secrecy outage probability than the traditional scheme without such a relay in the high $\rho^2$ regime for each case, which indicates that secrecy loss due to high correlation can be significantly reduced by using a cooperative jamming relay. However, the performance gains of our proposed scheme in the low $\rho^2$ regime are not obvious, possibly because the secrecy capacity for our proposed scheme is determined jointly by ${{\bf{h}}_{AE}}$ and ${{\bf{h}}_{RE}}$ and only their statistical information is known. By contrast, the secrecy capacity for the traditional scheme depends on only the statistical information of ${{\bf{h}}_{AE}}$. On the other hand, as \emph{M} increases, the control of the secrecy outage probability can be improve, which can be accounted for by the dependence of the Gamma distribution on \emph{M}.

\par Fig. 5 presents the secrecy outage probability$P_{out}$ versus the square of the correlation coefficient between the main and wiretap channels $\rho^2$ with ${\sigma _{RB}^2} = {\sigma _{RE}^2} = 1/2$ and ${\sigma _{RB}^2} = {\sigma _{RE}^2} = 1$, respectively. This simulation is conducted to verify that the shorter distance between Relay and Bob provides additional secrecy gains. Fig. 5 presents three cases, where the number of antennas at Alice is \emph{M} = 2, 4 and 8 and the number of antennas at Relay is \emph{N} = 2. The secrecy performance at larger ${\sigma _{RB}^2}$ and ${\sigma _{RE}^2}$ is better than that at smaller ${\sigma _{RB}^2}$ and ${\sigma _{RE}^2}$ because larger ${\sigma _{RB}^2}$ and ${\sigma _{RE}^2}$ represent a shorter distance and smaller path loss, which help to ensure that power is used more effectively.

\begin{figure}
\centering
\includegraphics[width=3.6in]{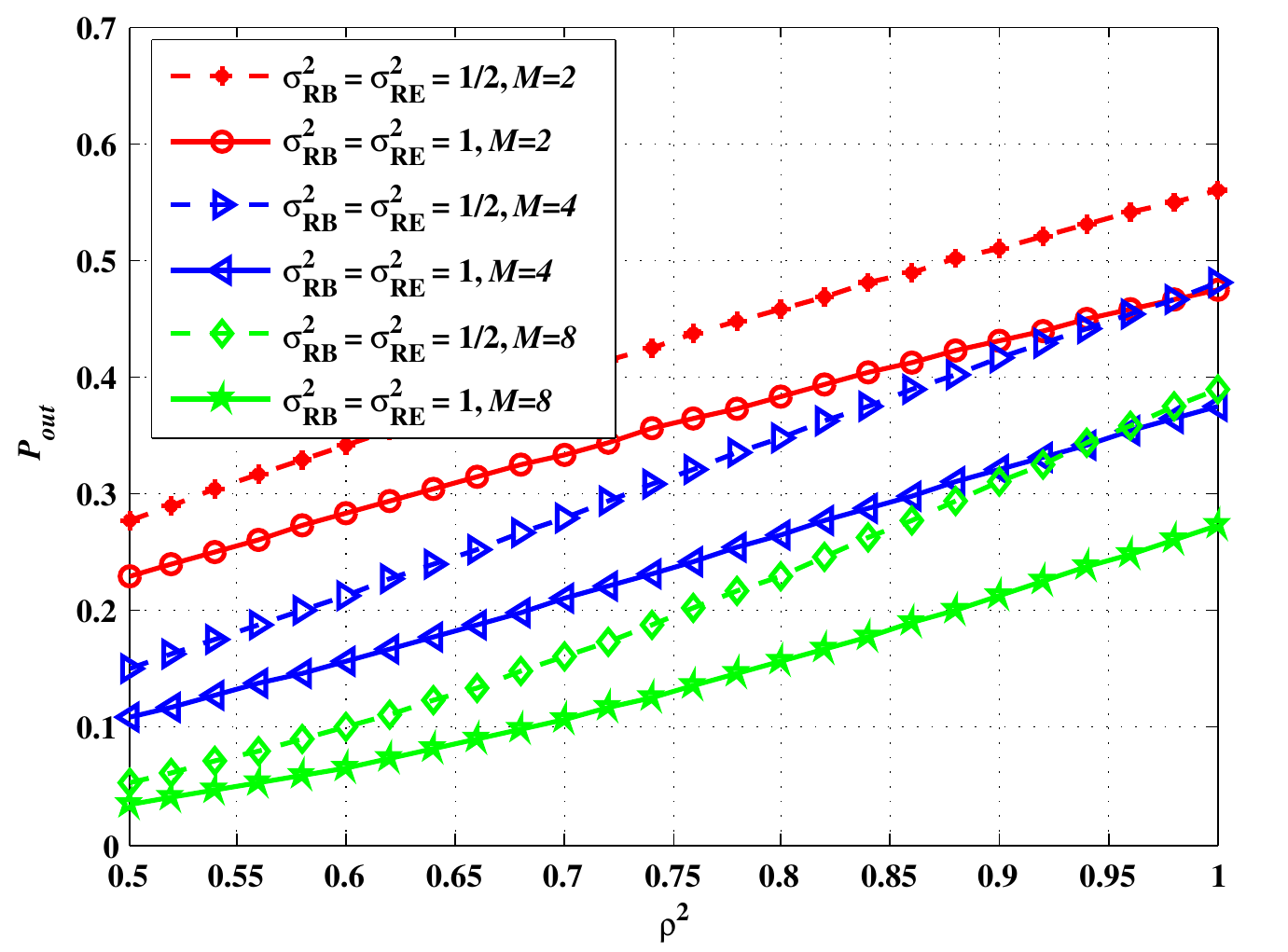}
\caption{The secrecy outage probability $P_{out}$ versus the square of the correlation coefficient between the main and wiretap channels $\rho^2$ with ${\sigma _{RB}^2} = {\sigma _{RE}^2} = 1/2$ and ${\sigma _{RB}^2} = {\sigma _{RE}^2} = 1$, respectively.}
\label{Fig4}
\end{figure}

\par Fig. 6 shows the secrecy outage probability $P_{out}$ with respect to the square of the correlation coefficient between the main and wiretap channels $\rho^2$, where the number of antennas Relay is \emph{N} = 2, 4 and 8, and the number of antennas at Alice is \emph{M} = 8. Moreover, ${\sigma _{RB}^2} = {\sigma _{RE}^2} = {\sigma _{RB}^2} = {\sigma _{RE}^2} = 1/2$ to remove the secrecy gains due to less path loss. It is obvious that our proposed scheme using a cooperative jamming relay has a much lower $P_{out}$ than the traditional scheme in each case (\emph{N} = 2, 4 and 8). Compared to the cases of \emph{N} = 2 and 4, the value of $P_{out}$ is the lowest, with the same $\rho^2$, for \emph{N} = 8, which means more antennas at Relay improves the secrecy. On the other hand, the attained secrecy gains via more antennas equipped at Relay are more obvious at high $\rho^2$ than that at low $\rho^2$.

\begin{figure}
\centering
\includegraphics[width=3.6in]{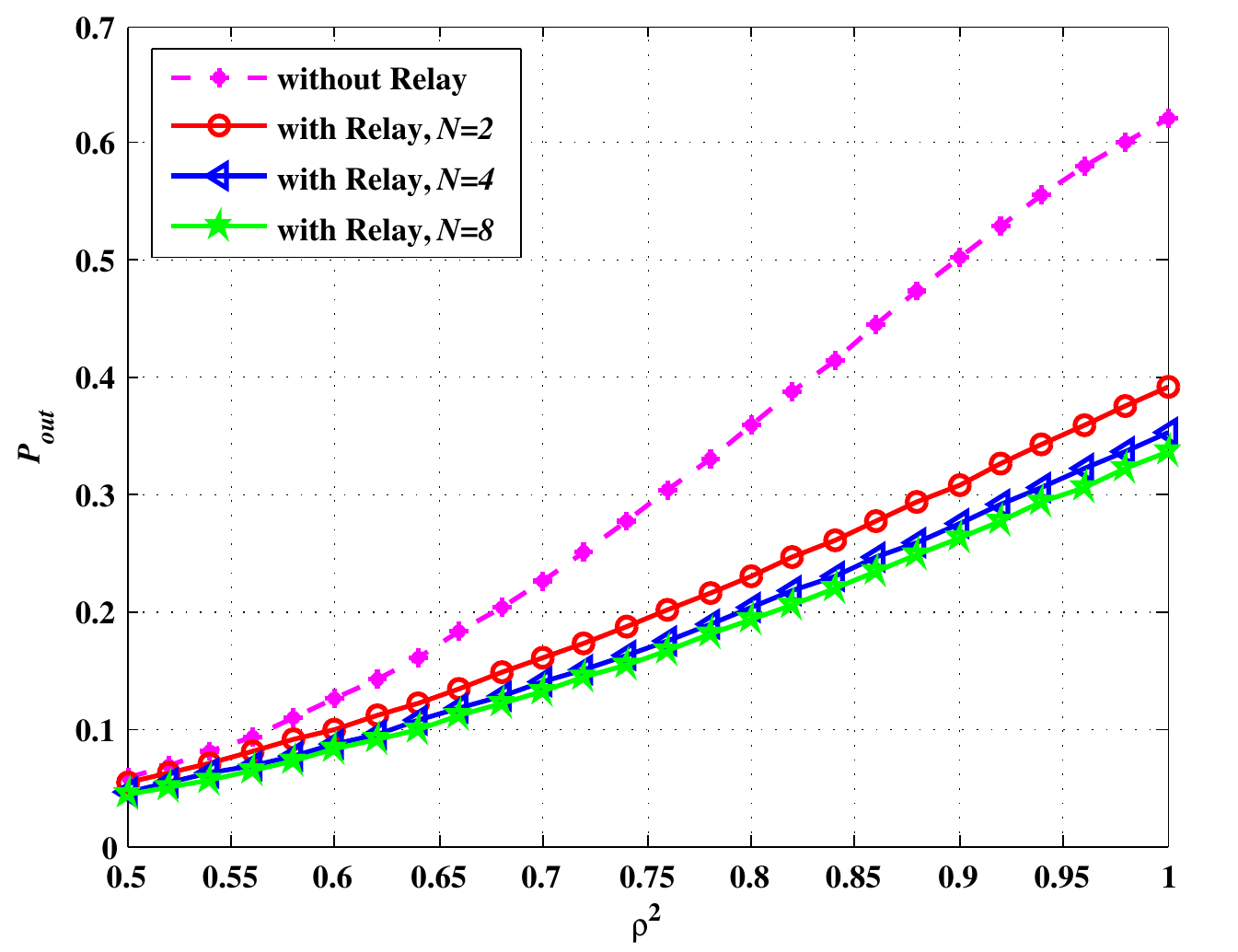}
\caption{The secrecy outage probability $P_{out}$ with respect to the square of the correlation coefficient between the main and wiretap channels $\rho^2$, where the number of antennas at Relay is \emph{N} = 2, 4 and 8, and the number of antennas at Alice is \emph{M} = 8.}
\label{Fig5}
\end{figure}

\subsection{$P_{out}$ versus $P$}

In this subsection, we discuss the relationship between the secrecy outage probability and the total transmitted power. Some of the parameters in the simulations are set as follows: the target secrecy rate $R_s$ is 0 bps/Hz; the additive noise at Bob and Eve is i.i.d. drawn from a complex Gaussian distribution with zero mean and unit variance; the variances of ${{\bf{h}}_{AB}}$, ${{\bf{h}}_{AE}}$, ${{\bf{h}}_{RB}}$ and ${{\bf{h}}_{RE}}$ are set to ${\sigma _{RB}^2} = {\sigma _{RE}^2} = {\sigma _{RB}^2} = {\sigma _{RE}^2} = 1/2$ to eliminate the secrecy gains due to less path loss.

\begin{figure}
\centering
\includegraphics[width=3.6in]{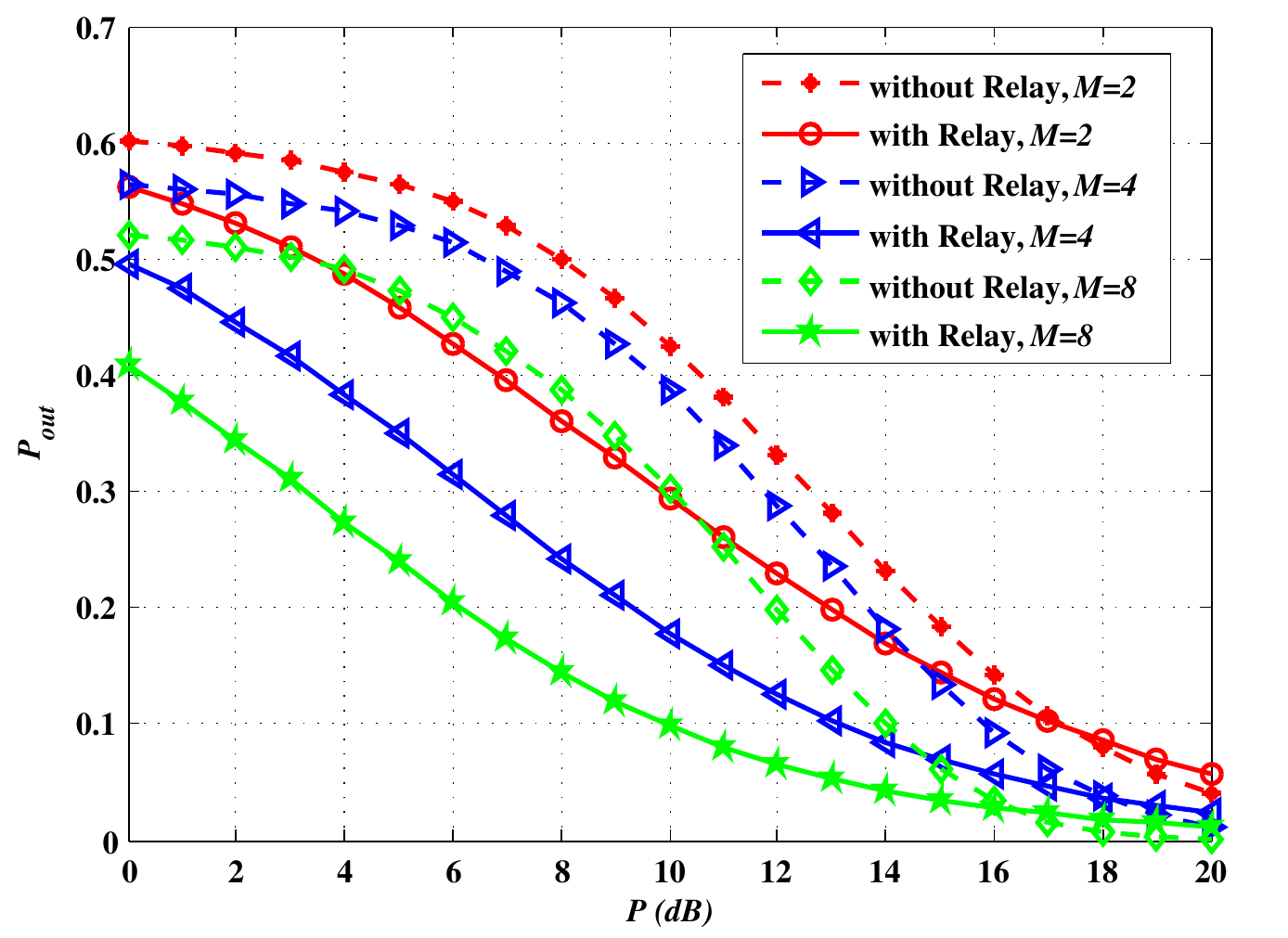}
\caption{The relationship between the secrecy outage probability $P_{out}$ and the total transmitted power $P(dB)$ in the case of $\rho^2 = 0.9$ and $R_s = 0 bps/Hz$.}
\label{Fig6}
\end{figure}

Fig. 7 plots the relationship between the secrecy outage probability $P_{out}$ and the total transmitted power $P(dB)$ in for $\rho^2 = 0.9$ and $R_s = 0 bps/Hz$. Fig. 8 presents three cases, where the number of antennas Alice \emph{M} = 2, 4 and 8 while the number of antennas at Relay is \emph{N} = 2. High power plays an important role in reducing $P_{out}$ and facilitating secrecy, which is consistent with our experience and knowledge. However, in the traditional scheme, huge $P$ is required to obtain low $P_{out}$. By contrast, the utilization of $P$ in our proposed scheme with a cooperative jamming relay is more efficient, which significantly decreases power consumption. In an extreme case of $\rho^2 = 1$, the secrecy outage probability in the traditional scheme remains almost unchanged as $P$ increases, and our proposed scheme has a distinct advantage, as shown in the Fig. 8. This result fully indicates that our proposed scheme of introducing a cooperative jamming relay plays an irreplaceable role in promoting secrecy at high $\rho^2$.

\begin{figure}
\centering
\includegraphics[width=3.6in]{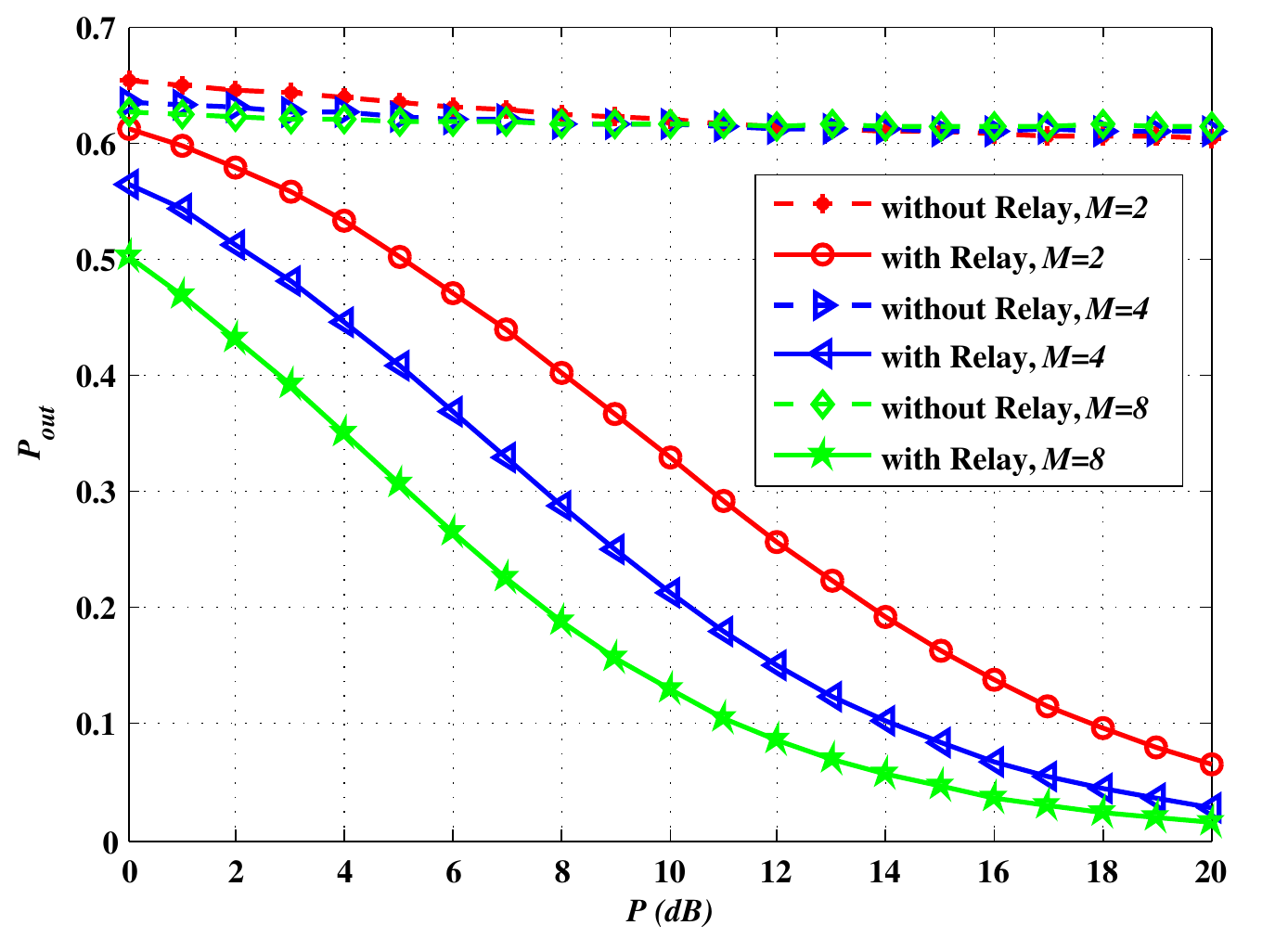}
\caption{The relationship between the secrecy outage probability $P_{out}$ and the total transmitted power $P(dB)$ in the case of $\rho^2 = 1$ and $R_s = 0 bps/Hz$.}
\label{Fig7}
\end{figure}

\par Fig. 9 shows the secrecy outage probability $P_{out}$ with respect to the total transmitted power $P(dB)$ in the case of $\rho^2 = 0.9$ and $R_s = 0 bps/Hz$, where the number of antennas Relay is \emph{N} = 2, 4 and 8 and the number of antennas at Alice is \emph{M} = 8. Our proposed scheme using a cooperative jamming relay has a much lower $P_{out}$ than that of the traditional scheme in each case (\emph{N} = 2, 4 and 8). Compared to the cases of \emph{N} = 2 and 4, the value of $P_{out}$ is the lowest, with the same $\rho^2$, when \emph{N} = 8, which means more antennas at Relay can help to improve the secrecy.

\begin{figure}
\centering
\includegraphics[width=3.6in]{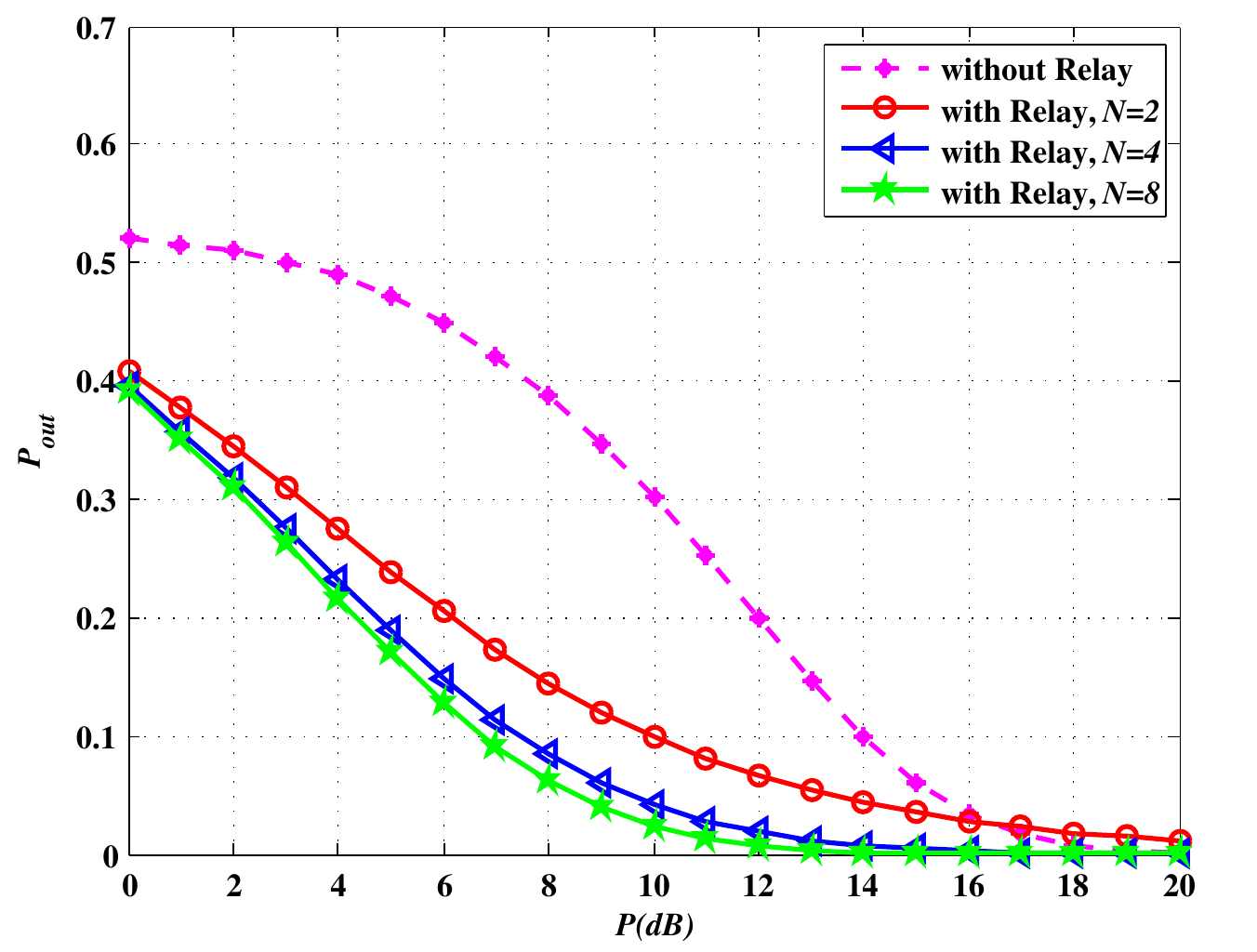}
\caption{The secrecy outage probability $P_{out}$ with respect to the total transmitted power $P(dB)$ in the case of $\rho^2 = 0.9$ and $R_s = 0 bps/Hz$.}
\label{Fig8}
\end{figure}

\subsection{$P_{out}$ versus $R_s$}

Here, we discuss the relationship between the secrecy outage probability and the target secrecy rate $R_s$. Some of the parameters in the simulations are set as follows: the total transmitted power \emph{P} is 3 dBW; the square of the correlation coefficient between the main and wiretap channels is $\rho^2 = 0.9$; the additive noise at Bob and Eve is assumed to be i.i.d. drawn from a complex Gaussian distribution with zero mean and unit variance; the variances of ${{\bf{h}}_{AB}}$, ${{\bf{h}}_{AE}}$, ${{\bf{h}}_{RB}}$ and ${{\bf{h}}_{RE}}$ are set to ${\sigma _{RB}^2} = {\sigma _{RE}^2} = {\sigma _{RB}^2} = {\sigma _{RE}^2} = 1/2$ to eliminate the secrecy gains due to path loss.

\begin{figure}
\centering
\includegraphics[width=3.6in]{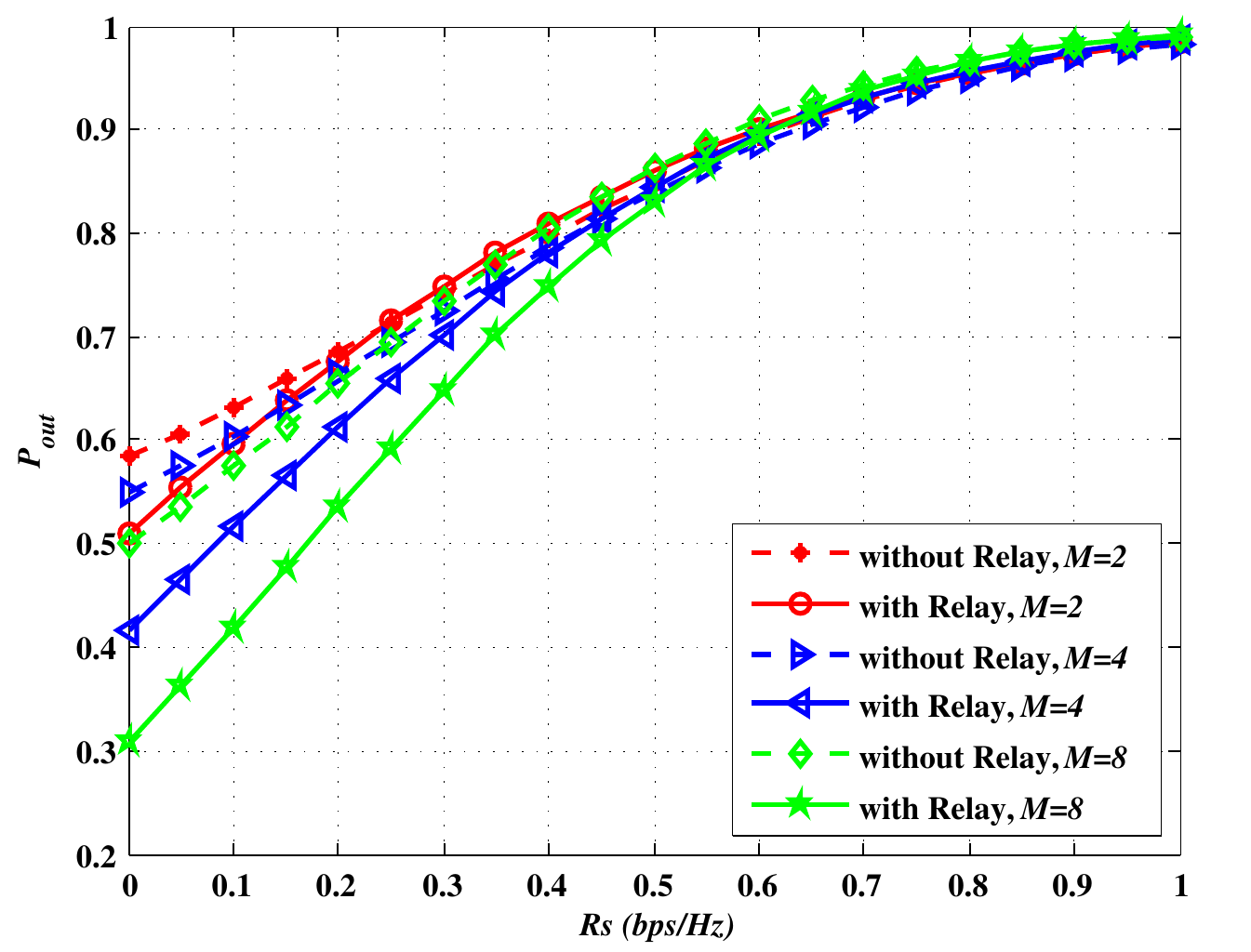}
\caption{The relationship between the secrecy outage probability $P_{out}$ and the target secrecy rate $R_s$ in the case of $\rho^2 = 0.9$ and $P = 3dB$.}
\label{Fig9}
\end{figure}

Fig. 10 plots the relationship between the secrecy outage probability $P_{out}$ and the target secrecy rate $R_s$ in the case of $\rho^2 = 0.9$ and $P = 3dB$. Fig. 10 presents three cases, where the number of antennas at Alice is \emph{M} = 2, 4 and 8 and the number of antennas at Relay is \emph{N} = 2. We observe that an increase in $R_s$
can rapidly aggravate the growth of $P_{out}$, which conforms to practical secrecy communications. When $R_s$ becomes very large, such as $R_s = 1$, $P_{out}$ approaches 1, which indicates that it is almost impossible to achieve secret communications. By comparison, our proposed scheme using a cooperative jamming relay has a lower $P_{out}$ than that of the traditional scheme in each case (\emph{M} = 2, 4 and 8), especially in the low $R_s$ regime. However, when $R_s$ is very large, both our proposed scheme and the traditional scheme have a very high $P_{out}$. By contrast, in the low $R_s$ regime, compared to the cases of \emph{M} = 2 and 4, the value of $P_{out}$ is the lowest, with the same $R_s$, for \emph{M} = 8. However, in the high $R_s$ regime, an increase in the number of antennas at Alice results in a high, rather than low, $P_{out}$.

\begin{figure}
\centering
\includegraphics[width=3.6in]{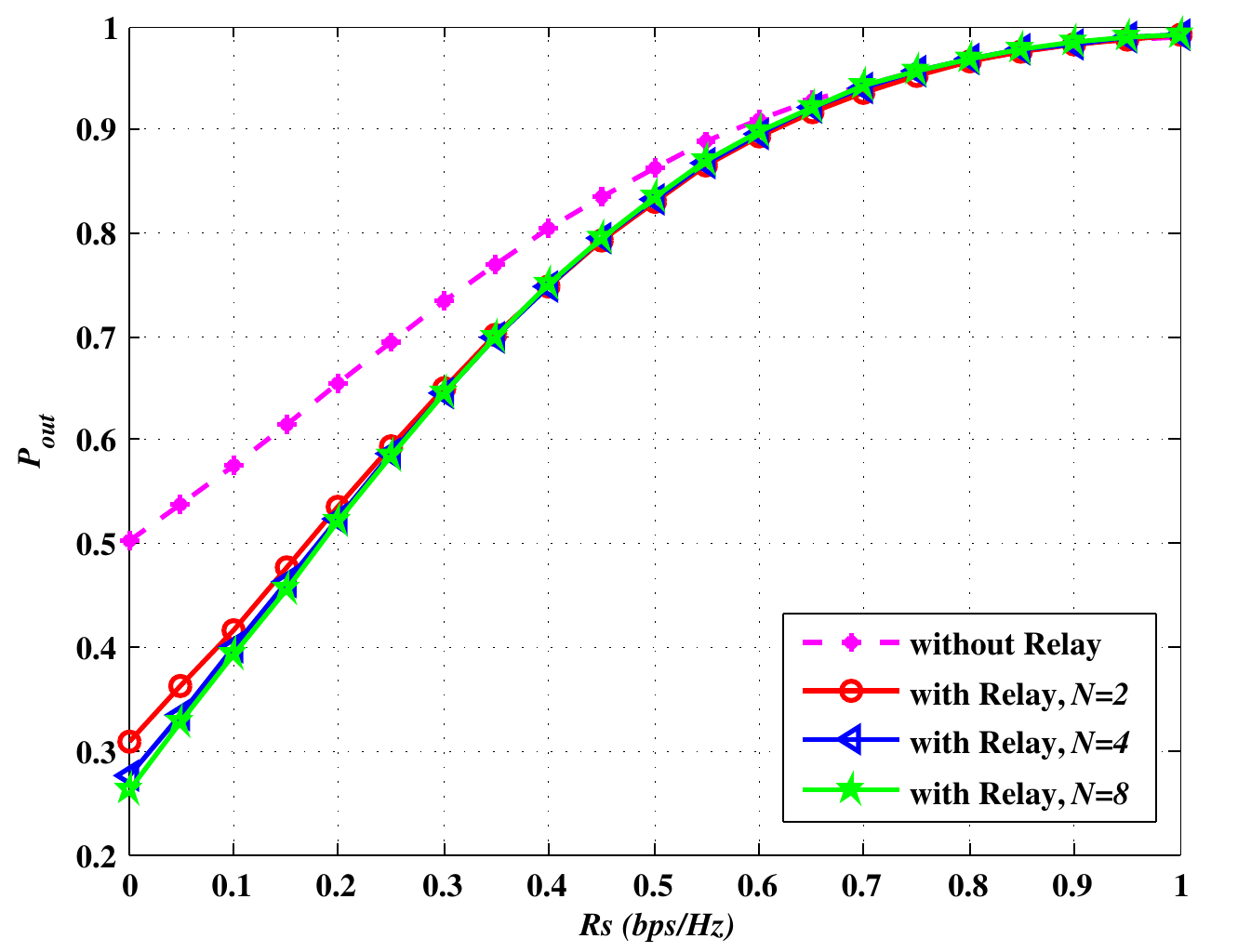}
\caption{The relationship between the secrecy outage probability $P_{out}$ and the target secrecy rate $R_s$ in the case of $\rho^2 = 0.9$ and $P = 3dB$.}
\label{Fig10}
\end{figure}

\par Fig. 11 shows the secrecy outage probability $P_{out}$ with respect to the target secrecy rate $R_s$ in the case of $\rho^2 = 0.9$ and $P = 3dB$, where the number of antennas at Relay is \emph{N} = 2, 4 and 8 and the number of antennas at Alice is \emph{M} = 8. Our proposed scheme using a cooperative jamming relay has a much lower $P_{out}$ than that of the traditional scheme, especially in the low $R_s$ regime. Three cases of \emph{N} = 2, 4 and 8 are considered in our proposed scheme. As shown in Fig. 11, the $P_{out}$ of the three cases in the same $R_s$ is similar. Therefore, the secrecy performance benefits little from additional antennas at Relay, and only a small gain is achieved at low $R_s$.

\section{CONCLUSIONS}

In this paper, we have investigated secrecy performance in the situation where the main and wiretap channels are highly correlated. Due to the high correlation, the secrecy suffers significant loss, which is difficult to effectively mitigate by employing signal processing techniques at Alice, such as precoding/beamforming and AN. In response to this challenging issue, we propose an ingenious scheme, in which a cooperative jamming relay is introduced, to facilitate confidential transmission. By reasonably deploying a cooperative jamming relay, the links from the relay to the involved legitimate receiver and eavesdropper can avoid high correlation. Then, AN is embedded into the null space of the channel from the relay to the legitimate receiver to degrade the reception at the eavesdropper; in this case, AN has been shown to be especially effective. The power allocation between the information-bearing signal and AN is investigated to achieve more efficient utilization of power and to maximize the secrecy. According to simulations, the traditional scheme of employing signal processing techniques at Alice suffers from high correlation more than our proposed scheme. Moreover, adding antennas at Alice and Relay facilitates secrecy at moderate total transmitted power and target secrecy rate. These results indicate that our ingeniously designed scheme of introducing a cooperative jamming relay achieves significant performance gain with respect to the secrecy outage probability. Note that although the cooperative jamming relay helps to achieve better secrecy, additional equipment is required and the complexity of the system design is increased.

% or


\ifCLASSOPTIONcaptionsoff
  \newpage
\fi


\begin{thebibliography}{1}

\bibitem{S1-1}
Bi, Suzhi, et al., ``Wireless communications in the era of big data,''\emph{IEEE Communications Magazine}, vol. 53, no. 10, pp. 190-199, Oct. 2015.

\bibitem{S1-2}
Liu Y, Chen H H and Wang L., ``Physical layer security for next generation wireless networks: theories, technologies, and challenges,'' \emph{IEEE Communications Surveys \& Tutorials}, Aug. 2016.

\bibitem{S1-3}
B. Schneier, ``Description of a new variable-length key, 64-bit block cipher (Blowfish),'' vol. 809, pp. 191-204, 1994

\bibitem{S1-4}
Hong Y W P, Kuo C C J. ``Enhancing physical-layer secrecy in multiantenna wireless systems: An overview of signal processing approaches,'' \emph{IEEE Signal Processing Magazine}, vol. 30, no. 5, pp. 29-40, 2013.

\bibitem{S1-5}
W. C.-Y. Lee, ``Effects on correlation between two mobile radio base-station antennas,'' \emph{IEEE Trans. Commun.}, vol. COM-21, pp.1214-1224, Nov. 1973.

\bibitem{S1-6}
 S. B. Rhee and G. I. Zysman, ``Results of suburban base-station spatial diversity measurements on the UHF bands,'' \emph{IEEE Trans. Commun.}, vol. COM-22, pp. 1630-1634, Oct. 1974.

\bibitem{S1-7}
 D.-S. Shiu, G. J. Foschini, M. J. Gans, and J. M. Kahn, ``Fading correlation and its effect on the capacity of multielement antenna systems,'' \emph{IEEE Trans. Commun.}, vol. 48, pp. 502-513, Mar. 2000.


\bibitem{S1-8}
 L. Dong, Z. Han, A. Petropulu, and H. Poor, ``Improving wireless physical layer security via cooperating relays,'' \emph{IEEE Trans. Signal Processing}, vol. 58, no. 3, pp. 1875-1888, Mar. 2010.

\bibitem{S1-9}
 X. He and A. Yener, ``Cooperation with an untrusted relay: A secrecy perspective,'' \emph{IEEE Trans. Inform. Theory}, vol. 56, no. 8, pp. 3807-3827, Aug. 2010.

\bibitem{S1-10}
 C. Jeong, I.-M. Kim, and D. I. Kim, ``Joint secure beamforming design at the source and the relay for an amplify-and-forward MIMO untrusted relay system,'' \emph{IEEE Trans. Signal Processing}, vol. 60, no. 1, pp. 310-325, Jan. 2012.

\bibitem{S1-11}
 Z. Ding, M. Peng, and H.-H. Chen, ``A general relaying transmission protocol for MIMO secrecy communications,'' \emph{ IEEE Trans. Commun.}, vol. 60, no. 11,pp. 3461-3471, Nov. 2012.

\bibitem{S1-12}
 J. Huang and A. L. Swindlehurst, ``Cooperative jamming for secure communications in MIMO relay networks,'' \emph{IEEE Trans. Signal Processing}, vol. 59, no. 10,pp. 4871-4884, Oct. 2011.

\bibitem{S1-13}
L. Dong, Z. Han, A. Petropulu, and H. Poor, ``Improving wireless physical layer security via cooperating relays,'' \emph{IEEE Trans. Signal Processing}, vol. 58, no. 3,pp. 1875-1888, Mar. 2010.

\bibitem{S1-14}
J. Li, A. Petropulu, and S. Weber, ``On cooperative relaying schemes for wireless physical layer security,'' \emph{IEEE Trans. Signal Processing}, vol. 59, no. 10,pp. 4985-4997, Oct. 2011.

\bibitem{S1-15}
G. Zheng, L.-C. Choo, and K.-K. Wong, ``Optimal cooperative jamming to enhance physical layer security using relays,'' \emph{IEEE Trans. Signal Processing}, vol. 59,no. 3, pp. 1317-1322, Mar. 2011.

\bibitem{S1-16}
H. Alves, G. Brante, R. D. Souza, D. B. d. Costa, and M. Latva-aho, ``On the performance of secure full-duplex relaying under composite fading channels,'' \emph{IEEE Signal Process. Lett.}, vol. 22, no. 7, pp. 867-870, Jul. 2015.

\bibitem{S1-17}
J.-H. Lee, ``Full-duplex relay for enhancing physical layer security in multi-Hop relaying systems,'' \emph{IEEE Commun. Lett.}, vol. 22, no. 7, Jul. 2015.

\bibitem{S1-18}
 G. Chen, Y. Gong, P. Xiao, and J. Chambers, ``Physical layer network security in the full-duplex relay system,'' \emph{ IEEE Trans. Inf. Forensics Security}, vol. 10, no. 3, pp. 574-583, Mar. 2015.

\bibitem{S1-19}
Jeon H, Kim N, Choi J, et al. ``Bounds on secrecy capacity over correlated ergodic fading channels at high SNR,'' \emph{IEEE Transactions on Information Theory}, vol. 75, no. 4, pp. 1975-1983, 2011.


\bibitem{S3-1}
Goel S, Negi R. ``Guaranteeing secrecy using artificial noise,'' \emph{IEEE Transactions on Wireless Communications}, vol. 7, no. 6, pp. 2180-2189, 2008.
\bibitem{S3-2}
Zhou X, Mckay M R. ``Secure transmission with artificial noise over fading channels: Achievable rate and optimal power allocation,'' \emph{IEEE Transactions on Vehicular Technology}, vol. 59, no. 8, pp. 3831-3842, 2010.
\bibitem{S3-3}
Alouini, M. S, and A. J. Goldsmith. ``Capacity of Rayleigh fading channels under different adaptive transmission and diversity-combining techniques,'' \emph{IEEE Transactions on Vehicular Technology}, vol. 48, no. 4, pp. 1165-1181, 1998.







\end{thebibliography}
\end{document}